\documentclass[graphicx, 12pt]{article}
\usepackage{indentfirst}

\usepackage{caption}
\usepackage{ctex}
\usepackage[nottoc]{tocbibind}
\usepackage{amsmath, array, graphicx, bm, cases}
\usepackage{enumerate, subfigure}
\usepackage{booktabs}
\ctexset{abstractname=Abstract}
\ctexset{figurename=Figure}
\ctexset{bibname=References}
\ctexset{tablename=Table}

\begin{document}

\small
\hoffset=-1truecm
\voffset=-2truecm
\title{\bf The shadow and gamma-ray bursts of a Schwarzschild black hole in asymptotic safety}

\author{Yuxuan Shi\footnote{E-mail address: shiyx2280771974@gmail.com}, Hongbo Cheng\footnote
{E-mail address: hbcheng@ecust.edu.cn}\\
Department of Physics,\\ East China University of Science and
Technology,\\ Shanghai 200237, China\\
The Shanghai Key Laboratory of Astrophysics,\\Shanghai 200234,
China}

\date{}
\maketitle

\begin{abstract}
We research on the neutrino pair annihilation $\nu+\overline{\nu}\longrightarrow e^{-}+e^{+}$ around a massive source in asymptotic safety. Since neutrinos and photons have the same geodesic equation around black holes, we can estimate the radius where the neutrinos will be released by obtaining a series of trajectory curves with various correction values $\xi$. The black hole shadow radius is influenced by the correction parameter $\xi$. The black hole shadow radius decreases with increasing the $\xi$. The ratio $\dfrac{\dot{Q}}{\dot{Q}_{Newt}}$ corresponding to the energy deposition per unit time over that in the Newtonian case is derived and calculated. We find that the quantum corrections to the black hole spacetime affect the emitted energy rate ratio for the annihilation. It is interesting that the more considerable quantum effect reduces the ratio value slightly. Although the energy conversion is damped because of the quantum correction, the energy deposition rate is enough during the neutrino-antineutrino annihilation. The corrected annihilation process can become a source of gamma ray burst. We also investigate the derivative $\dfrac{\mathrm{d}\dot{Q}}{\mathrm{d}r}$ relating to the star's radius $r$ to show that the quantum effect for the black hole will drop the ratio. The more manifest quantum gravity influence leads the weaker neutrino pair annihilation.
\end{abstract}

\vspace{3cm} \hspace{0cm} PACS number(s): 

\section{Introduction}

A lot of efforts from the astrophysical community have been
contributed to explain the gamma-ray bursts (GRBs) phenomenon with
enough energy source. GRBs are abrupt increases in gamma-ray
intensity coming from an identifiable point in the sky. After the
Big Bang, it is the most powerful celestial explosion.
Investigating the GRBs' energy source has been a focus of the
astrophysics community. One suitable theory is that the heated
accretion disk can annihilate neutrinos and antineutrinos into
electrons and positrons, $\nu\bar{\nu}\longrightarrow e^{-}e^{+}$.
This mechanism has the potential to produce GRBs energy [1-14]. A
newborn stellar-mass black hole that is accreting matter at
supercritical rates powers this system. Originating from the
standpoint of the annihilation process, high-energy gamma rays are
released when electron-positron couples close to the neutron
star's surface collapse, resulting in a supernova explosion. There
is an obvious explosion like this one. The efficiency of
annihilation of neutrino-antineutrino couples into
electron-positron pairs is more than 4 times that of the Newtonian
case in the background structure of a type II supernova, as
demonstrated by Ref [15]. This rate of annihilation increases by
30 times on the surface of a collapsing neutron star. A variety of
black holes that have been expanded upon by general relativity
have been examined recently. For a Kerr black hole with a thin
accretion disk, Ref [16, 17] examined the relativistic impact of
the annihilation process $\nu\bar{\nu}\longrightarrow e^{-}e^{+}$
on the energy deposition rate close to the rotation axis; Ref
[18, 19] addressed the implications of a strong off-axis.
Additional generalizations of general relativity, including higher
derivative gravity, Einstein dilation Gauss-Bonnet, charged
Galileon, Brans-Dicke, Born-Infeld generalization of
Reissner-Nordstrom solution, and Eddington-inspired Born-Infeld,
all dramatically increase the energy deposition rate in neutron
star and supernova envelopes [20]. When the quintessence field
was taken into account, the authors in Ref [21] found that
quintessence strengthened the energy of gamma ray bursts, which
could perhaps explain how gamma ray bursts originate. The
annihilation process of neutrinos around a massive source with a
f(R) global magnetic monopole is investigated in this study, and
it is found that this sort of black hole can be an even superior
supplier of gamma ray bursts than the beforehand kind of
explanation [22].

The black holes attract a lot of attentions from physical
community for several decades. As solutions to Einstein's field
equation in the general relativity, the black holes are also the
end-results of some stars during the process of gravitational
collapse [23-25]. In the case of black holes, the classical
description of spacetime can not be used to define the surrounding
of singularities inside the event horizon [23-25]. In order to
overcome the plague from the singular background, it is necessary
to generalize the general relativity up to the high energy scale
[26-35]. The generalization can lead the quantum gravity which
could resolve the unphysical black hole singularities [26-35].
Several proposals such as loop quantum gravity [26-30],
Horava-Lifshitz gravity [31-35], etc. have been put forward. Here
we are going to focus on the powerful issue named as asymptotic
safety scenario (AS) which is utilized to get rid of the
singularities of the black holes [35]. The techniques of the
functional renormalization group (FRG) was hired to limit the
results of the theory at trans-Planckian energies, making the
physical quantities to avoid the divergences at all scales [35].
The quantum effects are always shown as corrections to the
classical metrics [36-44]. A series of corrected metrics include
Schwarzschild [36], Kerr [37], Schwarzschild-AdS [38], Kerr-AdS
[39] and Reissner-Nordstrom-AdS [40-44] without the classical
singularity. More attentions have been contributed to the
RGI-Schwarzschild black holes. The evaporation of the
RGI-Schwarzschild black hole was investigated [45, 46], so was the
mini-black hole production in colliders [47]. The radial accretion
of matter onto a RGI-Schwarzschild black hole was also explored
[48-50]. The shadows of black holes have been studied in the AS
scenario for quantum gravity [51-54]. Recently, the quantum
gravity effects on the radiation properties of a thin accretion
disk surrounding a RGI-Schwarzschild black hole within the frame
of IR-limit of the AS theory were discussed [55].

Here we will explore the energy deposition rate by the neutrino
pair annihilation process around a Schwarzschild black hole in
asymptotic safety. The metric of Schwarzschild black hole modified
by the quantum effects needs to be studied in different
directions. The motion of test particle around RGI-Schwarzschild
black hole was described [56]. The electromagnetic ray reflection
spectroscopy for the corrected black hole was also considered [25,
57-59]. The quantum gravity effects on radiation characteristics
of thin accretion disks around a renormalization group improved
Schwarzschild black hole were shown [55]. Besides the black hole
features mentioned above, it is important to probe the
annihilation energy deposition rate in the background of this kind
of black holes to show the implications of asymptotic safety
introducing the quantum corrections to the spacetime structures of
the gravitational sources. However, to the best of our knowledge,
little contributions have been paid for this topic. In this paper,
we derive the integral form of the neutrino pair annihilation
efficiency based on the Schwarzschild black hole with running
Newton coupling $G(r)$ corresponds to the quantum effects.
Secondly, we calculate the ratio of total energy deposition to
total Newtonian energy deposition for parameter $\xi$ encoding the
quantum effects on the spacetime geometry With the help of our
numerical estimation, we wil discuss the influence from a
RGI-Schwarzschild black hole in the IR-limit of the asymptotic
safety scenario on the possibility that the astrophysical bodies
attract the annihilation process generating the gamma-ray burst.
The results and conclusions are listed in the end.

\section{The effective potential and photon orbits of the RGI-Schwarzschild
black hole}

We adopt the Schwarzschild metric modified by the quantum gravity in the infrared
limit as follow [60-64],
\begin{align}
\mathrm{d}s^{2}=f(r)\mathrm{d}t^{2}-\dfrac{\mathrm{d}r^{2}}{f(r)}-r^{2}\left(\mathrm{d}\theta^{2}+\sin^{2}\theta\mathrm{d}\varphi^{2}\right)
\end{align}
where
\begin{align}
f(r)=1-\dfrac{2MG(r)}{r}
\end{align}
with mass $M$. For the quantum-gravity-corrected
Schwarzschild metric in the infrared limit, the running coupling
$G(r)$ takes the form [61],
\begin{align}
G(r)=G_{0}\left(1-\dfrac{\xi}{r^{2}}\right)
\end{align}
where $\xi$ is a parameter with dimensions of length squared associated to the scale identification between the momentum scale and the radial coordinate. The general function of radial coordinate $G(r)$ as the generalization of gravitational constant, comes as a result of quantum effects. It should be pointed out that the running coupling $G(r)$ contains the parameter $\xi$ standing for the renormalization group improvement and the parameter will change the spacetime structure [36, 54]. There are some upper bounds on $\xi$ derived from a series of gravitational and nongravitational measurements [54]. We can choose $c=G_{0}=1$ in natural units. The radius of the event horizon is a root of $f(r)=0$ [55]. According to the calculation and discussion, the dimensionless parameter
$\overline{\xi}=\dfrac{\xi}{M^{2}}$ and the mass $M$ should be limited as [55],
\begin{align}
\begin{cases}
  0\leq\overline{\xi}\leq\overline{\xi}_{c} \\
  M>M_{c} \\
\end{cases}
\end{align}
where $\overline{\xi}_{c}=\dfrac{16}{27}$ and $M_{c}=\sqrt{\dfrac{27}{16}\xi}$, to keep the metric with two horizons [55]. It is significant to investigate the quantum gravity effects on the annihilation process within the region (4).

It is necessary to elaborate the accretion disk of Schwarzschild black holes in asymptotic safety because only sufficiently hot accretion disks can emit the neutrinos [65-68]. We plan to discuss the geodesics for a particle motion in the background specified by metric (1). We limit the particle motion to the equatorial plane with $\theta=\dfrac{\pi}{2}$ to write the Lagrangian as the description of the planar motion given by [15, 69],
\begin{align}
\mathcal{L}&=\dfrac{1}{2}g_{\mu\nu}\dfrac{\mathrm{d}x^{\mu}}{\mathrm{d}\tau}
\dfrac{\mathrm{d}x^{\nu}}{\mathrm{d}\tau}\hspace{3cm}\notag\\
&=\dfrac{1}{2}\left[f(r)\left(\dfrac{\mathrm{d}t}{\mathrm{d}\tau}\right)^{2}
-\dfrac{1}{f(r)}\left(\dfrac{\mathrm{d}r}{\mathrm{d}\tau}\right)^{2}
-r^{2}\left(\dfrac{\mathrm{d}\varphi}{\mathrm{d}\tau}\right)^{2}\right]
\end{align}
then the components of particle's momentum are,
\begin{align}
E&=f(r)\dfrac{\mathrm{d}t}{\mathrm{d}\tau}\\
L&=r^{2}\dfrac{\mathrm{d}\varphi}{\mathrm{d}\tau}
\end{align}
The geodesics introduce $E$ and $L$ are their own
constants and the momenta satisfy $P_{\mu}P^{\mu}=0$ for photons
[15]. The equation for radial motion is [15],
\begin{align}
\left(\dfrac{\mathrm{d}r}{\mathrm{d}\tau}\right)^{2}+V_{eff}(r)=\dfrac{1}{b^{2}}
\end{align}
where the effective potential is,
\begin{align}
V_{eff}(r)=\dfrac{1}{r^{2}}\left[1-\dfrac{2M}{r}\left(1-\dfrac{\xi}{r^{2}}\right)\right]
\end{align}
\begin{table}
\caption{The event horion $r_+$, the photon sphere $r_p$ and the impact parameter $b_p$ of the photon sphere with the different $\xi$.}
\centering
\begin{tabular}{lcccccc}
    \toprule
    \quad&$\xi=0$&$\xi=1/9$&$\xi=2/9$&$\xi=1/3$&$\xi=4/9$&$\xi=5/9$\\
    \midrule[1.5pt]
    $r_+$&$2.00000$&$1.94102$&$1.87336$&$1.79252$&$1.68806$&$1.51749$\\
    $r_p$&$3.00000$&$2.93553$&$2.86460$&$2.78514$&$2.69375$&$2.58397$\\
    $b_p$&$5.19615$&$5.12975$&$5.05818$&$4.98013$&$4.89356$&$4.79502$\\
    \toprule
\end{tabular}
\end{table}
\begin{figure}[t]
\centering
\subfigure[$\xi=1/3$]{\includegraphics[width=8cm]{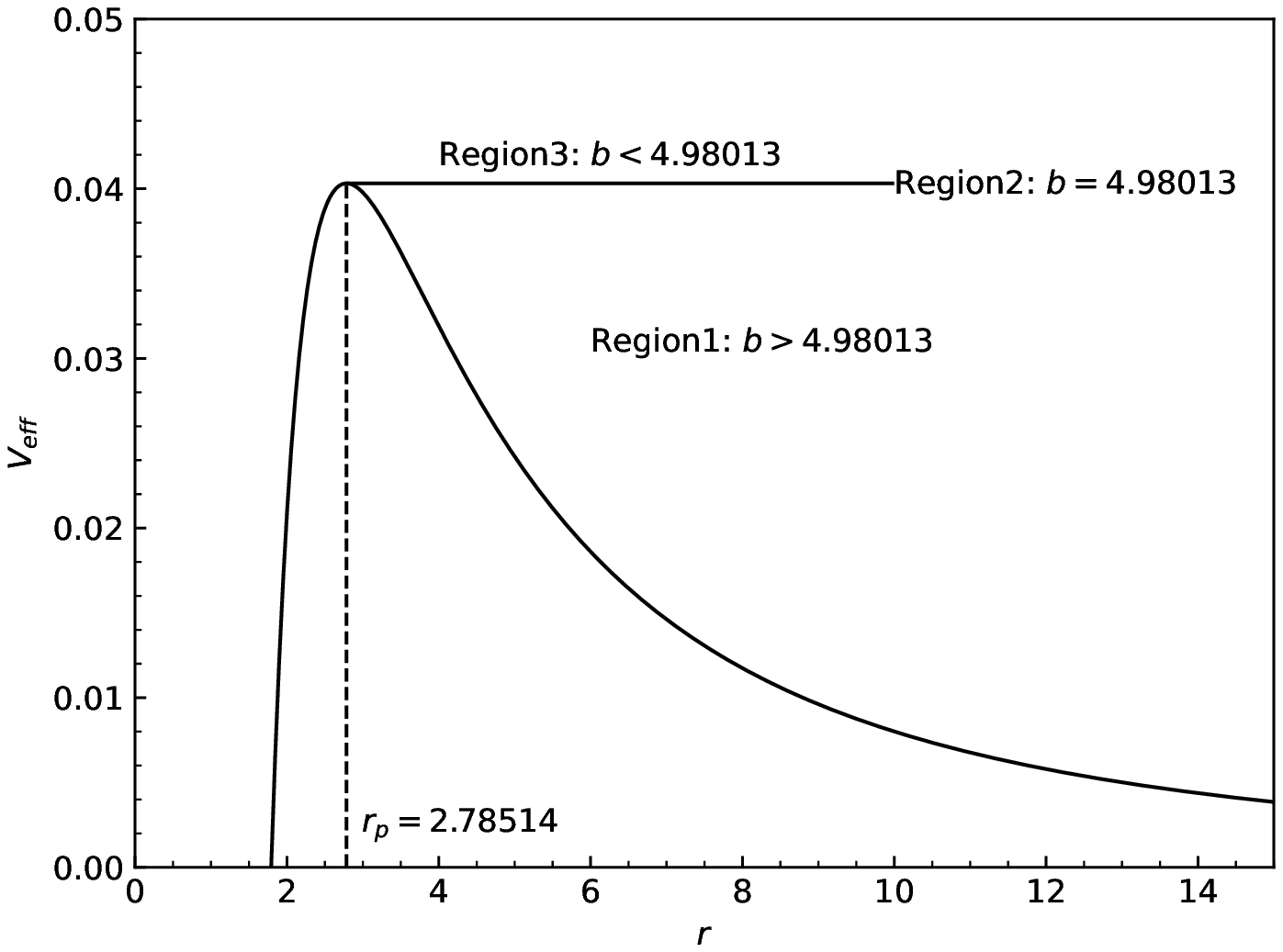}}
\subfigure[$\xi=5/9$]{\includegraphics[width=8cm]{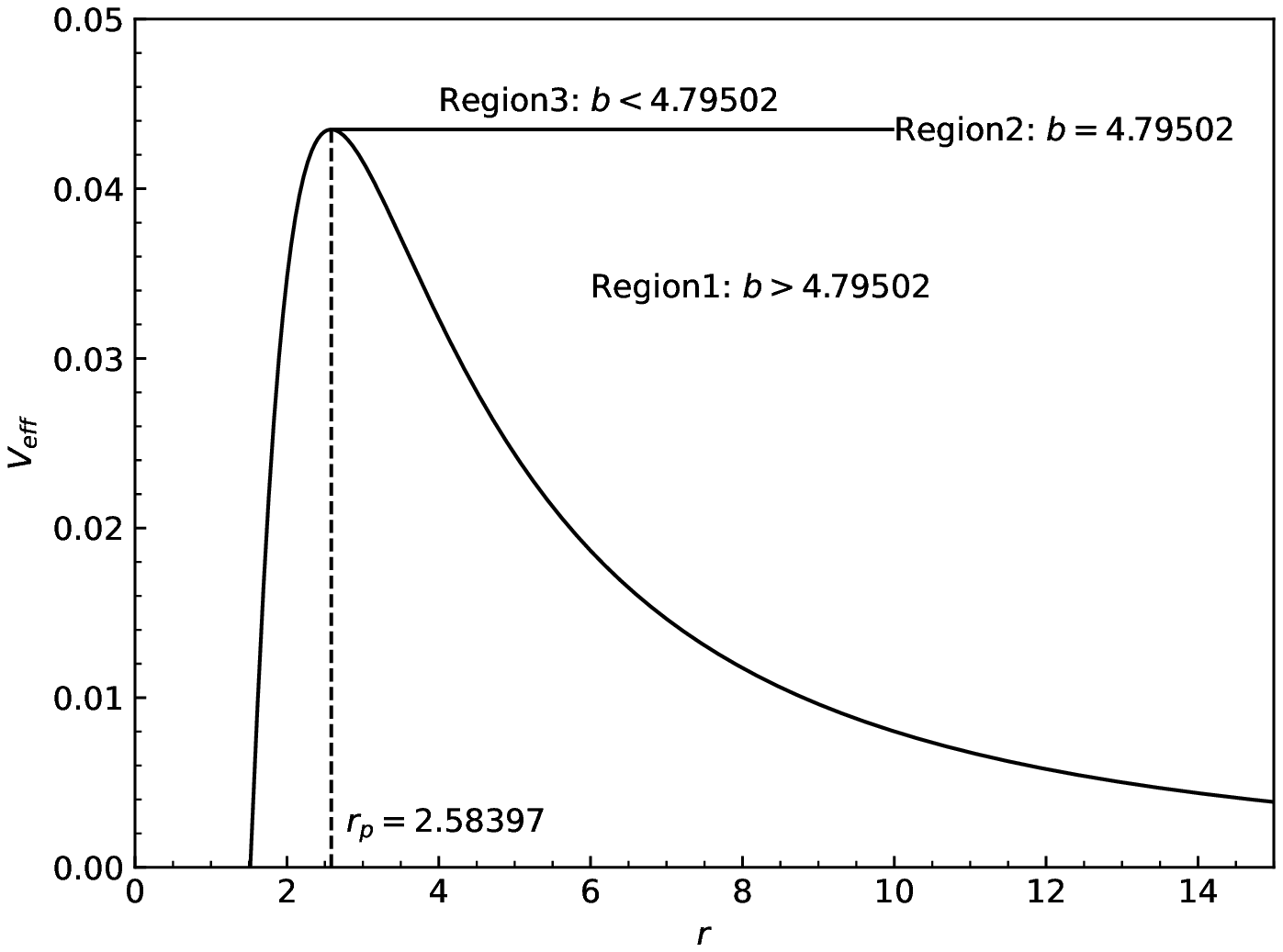}}
\caption{The effective potential profiles with $M = 1$ for $\xi = \dfrac{1}{3}$ (left panel) and $\xi = \dfrac{5}{9}$ (right panel). The photon sphere $r_{p}$ has radii indicated by dashed lines.}
\end{figure}
Here the impact factor is defined as $b=\dfrac{L}{E}$. In view of the conditions of photon sphere like $\dfrac{\mathrm{d}r}{\mathrm{d}\tau}=0$ and $\dfrac{\mathrm{d}^{2}r}{\mathrm{d}\tau^{2}}=0$, Eq.(9) leads,
\begin{align}
V_{eff}(r)&=\dfrac{1}{b^{2}}\\
\dfrac{\mathrm{d}V_{eff}}{\mathrm{d}r}&=0
\end{align}
or equivalently,
\begin{align}
r_{p}^{2}&=b_{p}^{2}f(r_{p})\\
r_{p}^{3}\dfrac{\mathrm{d}f(r)}{\mathrm{d}r}\biggl|_{r=r_{p}}&=2b_{p}^{2}f^{2}\left(r_{p}\right)
\end{align}
where $r_{p}$ is the radius of photon sphere and $b_{p}$ is the relevant impact factor. According to the discussions on the metric (1), the outer horizon $r_{+}$, radius of photon sphere $r_{p}$ and impact factor $b_{p}$ relate to the dimensionless parameter $\xi$ associated with the quantum gravity and the relations are listed in the Table 1. We plot the effective potential as function of radial coordinate in Figure 1. These figures show that the quantum gravity effect amends the features of the compact source although the shapes of curves corresponding to the different values of parameter $\xi$ are similar.By combining the Eq.(7), Eq.(8) and Eq.(9), we have the trajectory equation like [15],
\begin{align}
\dfrac{\mathrm{d}r}{\mathrm{d}\varphi}=\pm
r^{2}\sqrt{\dfrac{1}{b^{2}}-\dfrac{1}{r^{2}}\left[1-\dfrac{2M}{r}
\left(1-\dfrac{\xi}{r^{2}}\right)\right]}
\end{align}
By setting $u\equiv\dfrac{1}{r}$, we can transform (14) into
\begin{align}
\dfrac{\mathrm{d}u}{\mathrm{d}\varphi}=\sqrt{\dfrac{1}{b^2} - u^2\left[1-2Mu\left(1-\xi u^2\right)\right]}\equiv G(u)
\end{align}

For the sake of distinguishing among the light trajectories, it is
useful to establish orbital fractions $n=\dfrac{\varphi}{2\pi}$
[15, 69]. The direct emission of these rays corresponds to
$n<\dfrac{3}{4}$, the lensing ring to
$\dfrac{3}{4}<n<\dfrac{5}{4}$, and the photon ring to
$n>\dfrac{5}{4}$. These rays intersect the equatorial plane once,
twice, and more than twice. The range of the incident parameter
$b$ of the photon ring, lens ring, and direct image for the
situations with $\xi=\dfrac{1}{3}$ and $\xi=\dfrac{5}{9}$ is given
below,
\begin{align}
\xi = \dfrac{1}{3}\Rightarrow
\begin{cases}
    \text{Direct emission:}n<\dfrac{3}{4},\quad&b<4.74314\quad\text{and}\quad b>6.04424\\
    \text{Lensing ring:}\dfrac{3}{4}<n<\dfrac{5}{4},\quad&4.74314<b<4.96569\quad\text{and}\quad5.02321<b<6.04424\\
    \text{Photon ring:}n>\dfrac{5}{4},\quad&4.96569<b<5.02321
\end{cases}
\end{align}
\begin{align}
\xi = \dfrac{5}{9}\Rightarrow
\begin{cases}
    \text{Direct emission:}n<\dfrac{3}{4},\quad&b<4.43019\quad\text{and}\quad b>5.95531\\
    \text{Lensing ring:}\dfrac{3}{4}<n<\dfrac{5}{4},\quad&4.43019<b<4.76382\quad\text{and}\quad4.85521<b<5.95531\\
    \text{Photon ring:}n>\dfrac{5}{4},\quad&4.76382<b<4.85521
\end{cases}
\end{align}
\begin{figure}[t]
\centering
\subfigure[$\xi=1/3$]{\includegraphics[width=8cm]{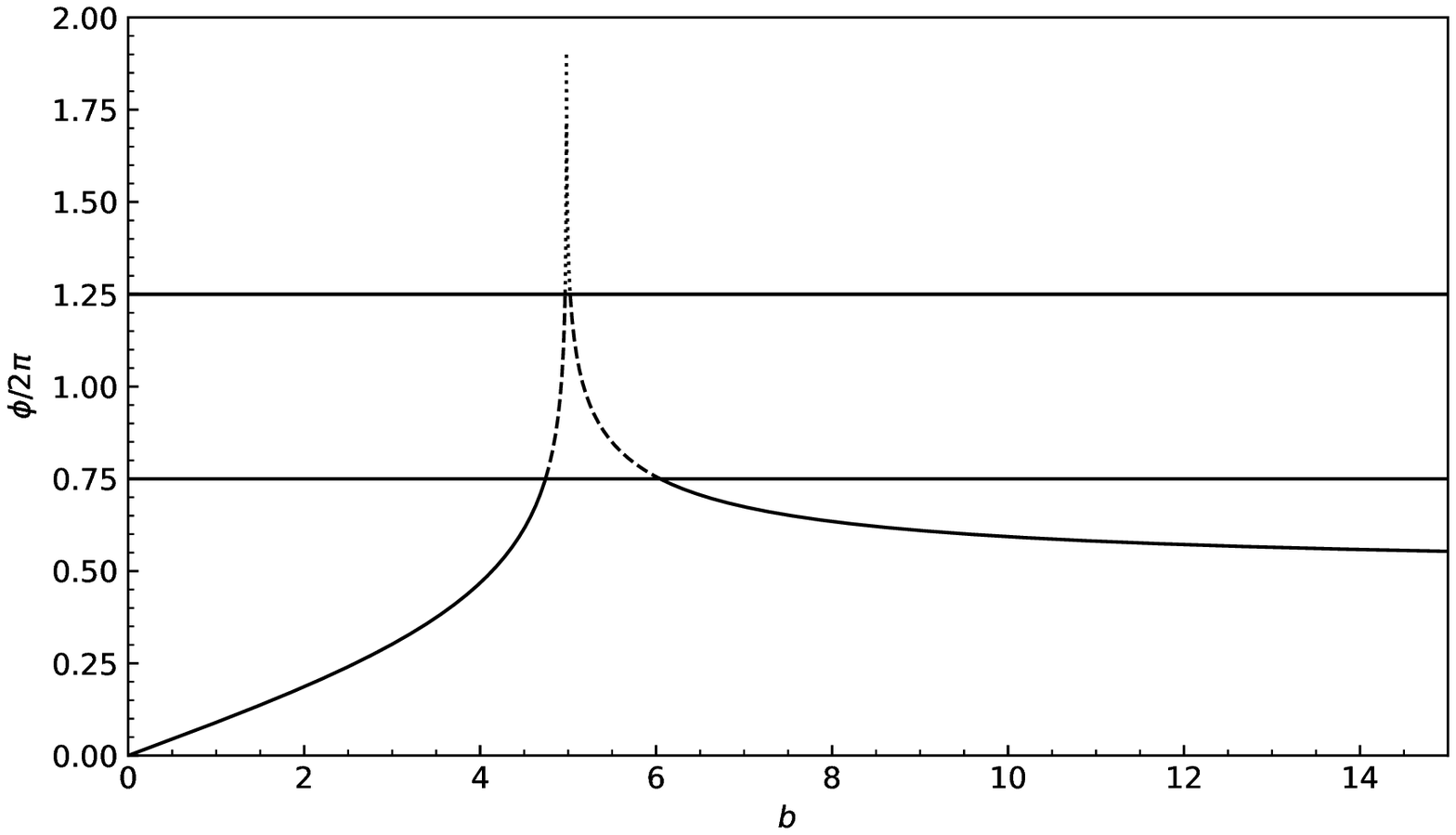}}
\subfigure[$\xi=5/9$]{\includegraphics[width=8cm]{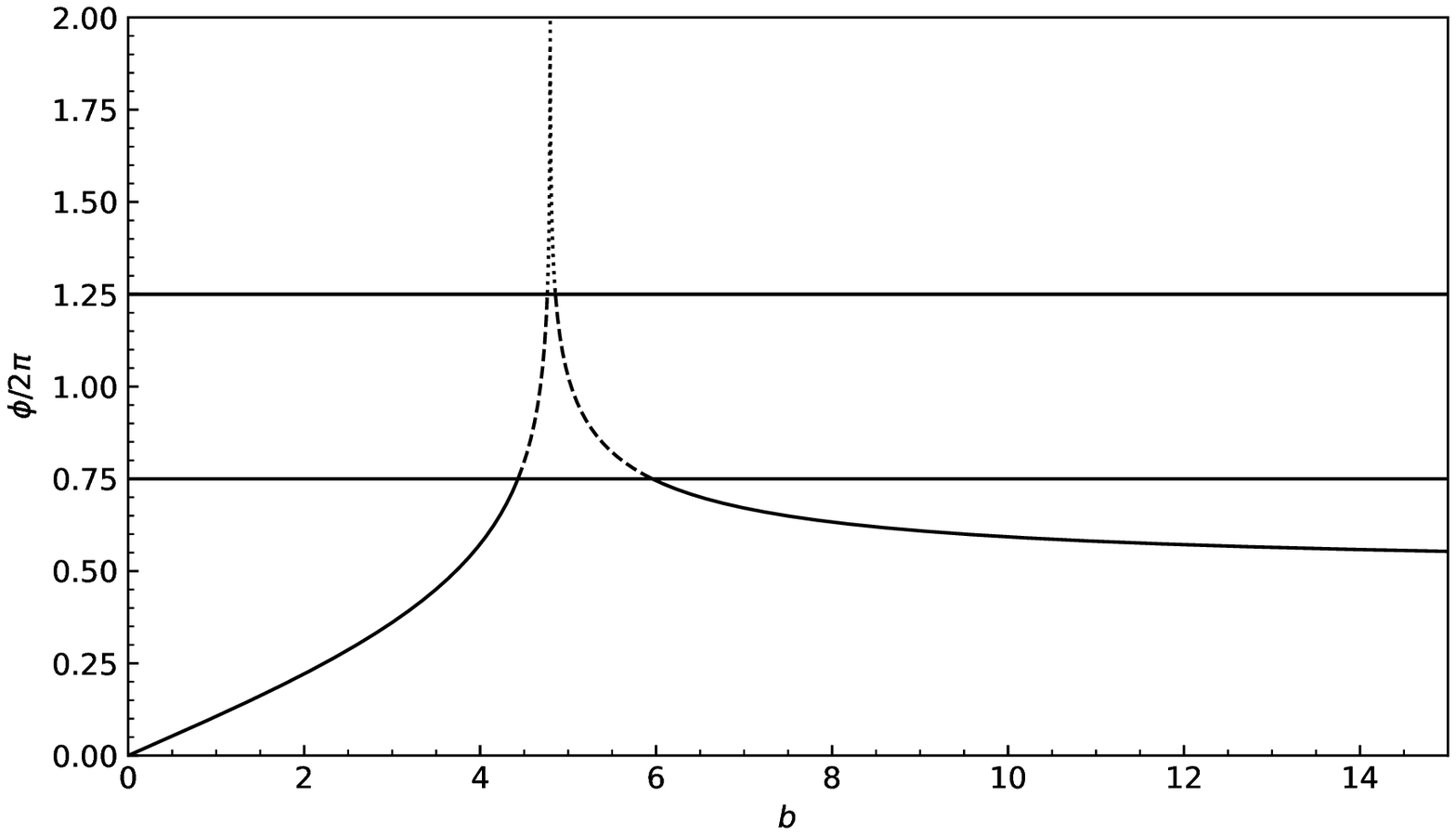}}\\
\subfigure[$\xi=1/3$]{\includegraphics[width=8cm]{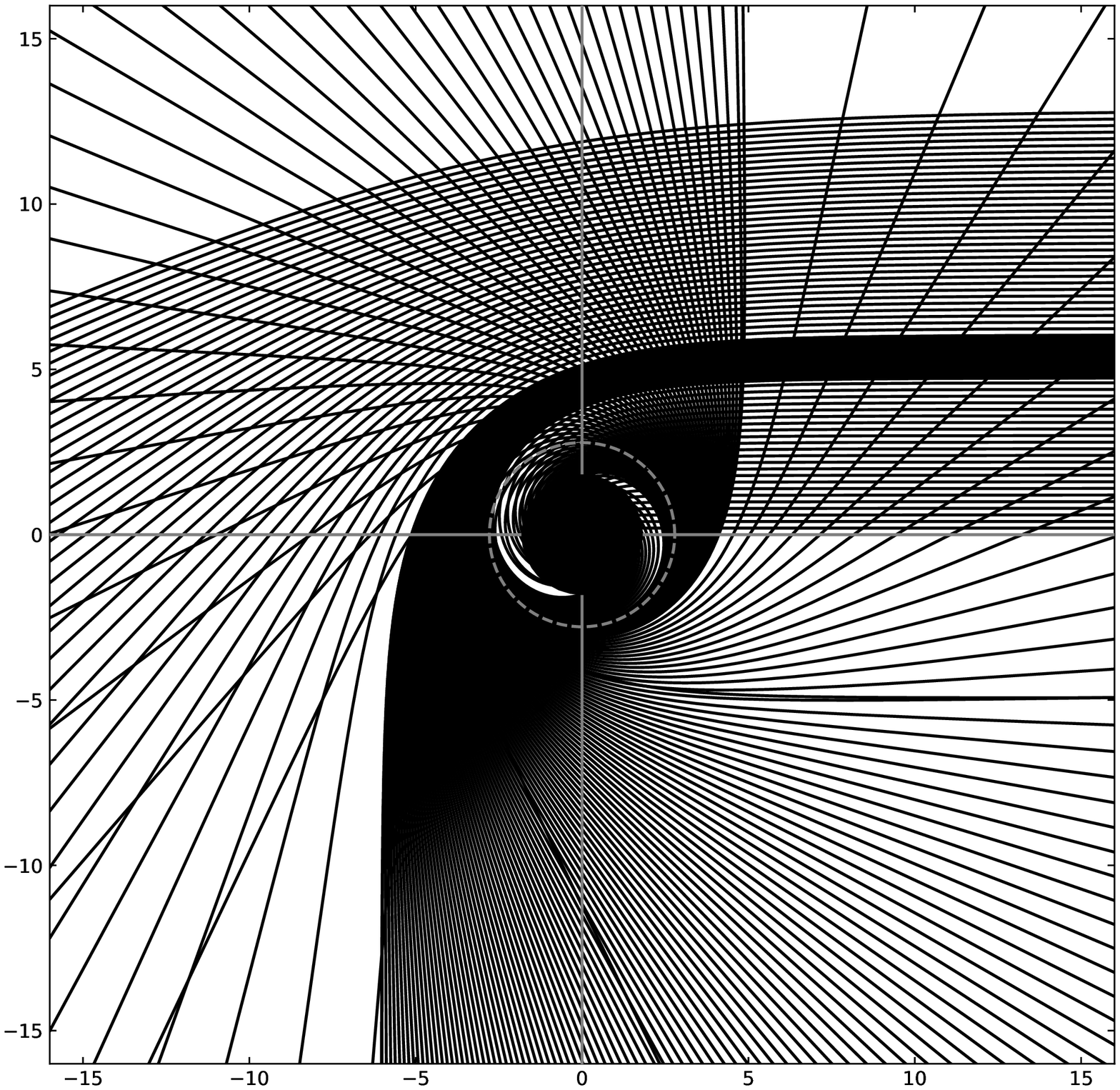}}
\subfigure[$\xi=5/9$]{\includegraphics[width=8cm]{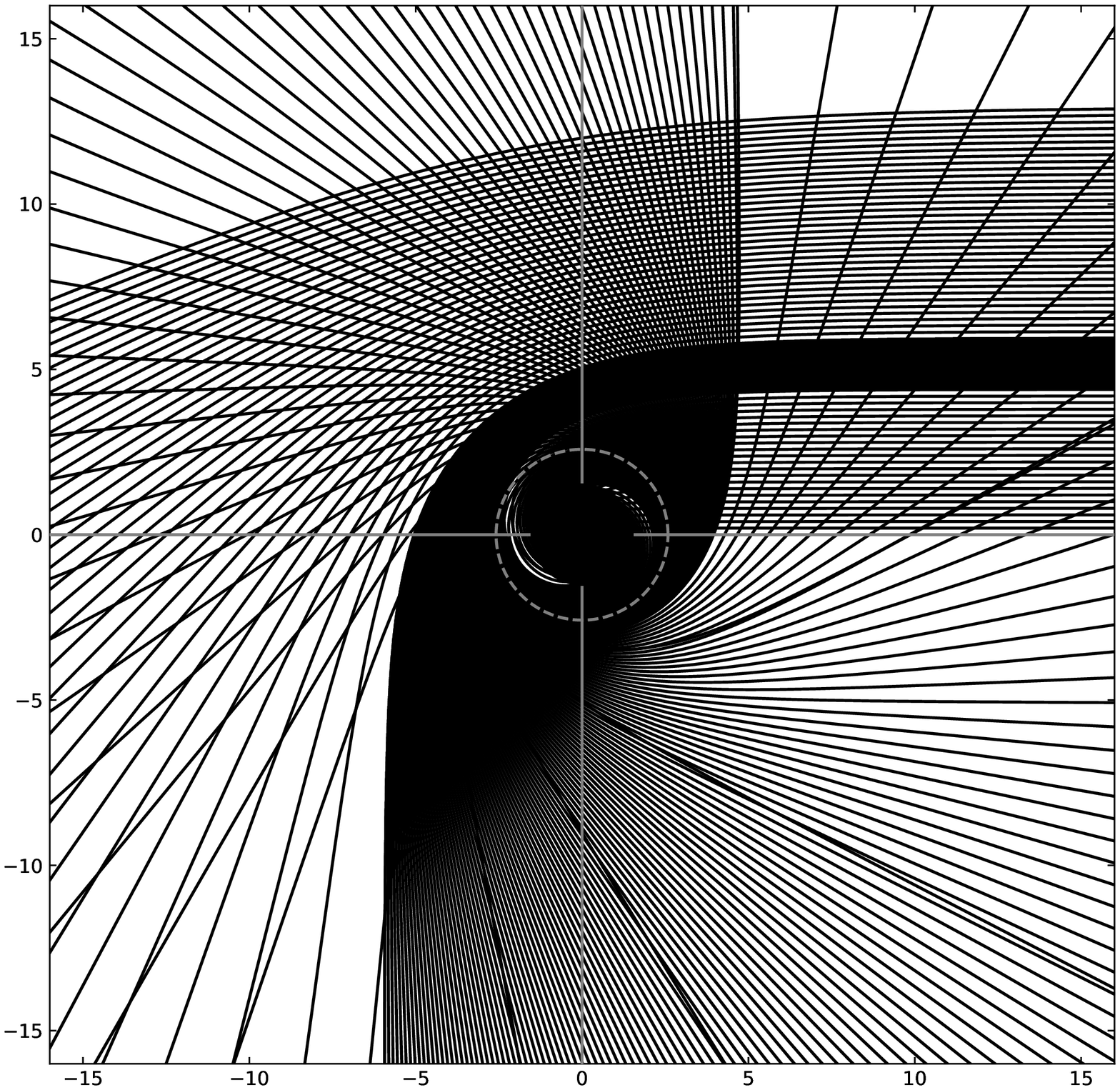}}
\caption{The way photons behave in the RGI-Schwarzschild spacetime depends on the impact parameter $b$.}
\end{figure}

The Figure 2 describes a series of photon trajectories with
coordinates $(r,\varphi)$ on the equatorial plane. In the Figure
2, The solid discs stand for black holes, and the photon orbit is
represented by the dashed grey line. When $b<b_{p}$, the
electromagnetic rays directly enters the black hole. The photons
orbit the gravitational source with the constraint $b=b_{p}$. The
potential shown in Figure 1 causes the light to deflect, and
photons pass through and escape from the black hole if the impact
parameter is larger, such as $b>b_{p}$. Some photons escape the
compact body and the remaining ones are either absorbed by the
body or go around the source if the light is directed towards the
black hole.

By comparing the aforementioned equations with those of the
RGI-Schwarzschild black hole, it is clear that the parameter $\xi$
decreases the regions of the direct emission, the lensing ring,
and the photon ring. The black hole's accretion disc is about to
form and develop. Neutrinos are released from the hotter disc.

\section{The energy deposition rate by the neutrino annihilation process}

We start to consider the energy deposition in the spacetime governed by Eq.(2) and (3).
The energy deposition per unit time and per unit volume for the neutrino pair annihilation process is given by [15],
\begin{align}
\dfrac{\mathrm{d}E(\mathbf{r})}{\mathrm{d}t\mathrm{d}V}=2KG_{F}^{2}F(r)\iint
n(\varepsilon_{\nu})n(\varepsilon_{\overline{\nu}})
(\varepsilon_{\nu}+\varepsilon_{\overline{\nu}})
\varepsilon_{\nu}^{3}\varepsilon_{\overline{\nu}}^{3}
\mathrm{d}\varepsilon_{\nu}\mathrm{d}\varepsilon_{\overline{\nu}}
\end{align}
where
\begin{align}
K=\dfrac{1}{6\pi}(1\pm4\sin^{2}\theta_{W}+8\sin^{4}\theta_{W})
\end{align}
with the Weinberg angle $\sin^{2}\theta_{W}=0.23$. For
various neutrino pairs, the forms of Eq.(6) can be chosen as [15],
\begin{align}
K(\nu_{\mu},\overline{\nu}_{\mu})=K(\nu_{\tau},\overline{\nu}_{\tau})
=\dfrac{1}{6\pi}\left(1-4\sin^{2}\theta_{W}+8\sin^{4}\theta_{W}\right)
\end{align}
and
\begin{align}
K(\nu_{e},\overline{\nu}_{e})
=\dfrac{1}{6\pi}\left(1+4\sin^{2}\theta_{W}+8\sin^{4}\theta_{W}\right)
\end{align}
respectively [15]. Here the Fermi constant $G_{F}=5.29\times 10^{-44}cm^{2}MeV^{-2}$.
The angular integration factor is represented by [15],
\begin{align}
F(r)&=\iint\left(1-\bm{\Omega_{\nu}}\cdot\bm{\Omega_{\overline{\nu}}}\right)^{2}
\mathrm{d}\Omega_{\nu}\mathrm{d}\Omega_{\overline{\nu}}\notag\\
&=\dfrac{2\pi^{2}}{3}(1-x)^{4}\left(x^{2}+4x+5\right)\hspace{1cm}
\end{align}
where
\begin{align}
x=\sin\theta_{r}
\end{align}
The angle $\theta_{r}$ is between the particle trajectory and the tangent vector
to a circle orbit at radius $r$. For one kind of neutrino and antineutrino,
$\Omega_{\nu}(\Omega_{\overline{\nu}})$ is the unit direction vector and
$\mathrm{d}\Omega_{\nu}(\mathrm{d}\Omega_{\overline{\nu}})$ is a solid angle.
At temperature $T$, $n(\varepsilon_{\nu})$ and $n(\varepsilon_{\overline{\nu}})$
are number densities for neutrino and antineutrino respectively in the phase space
and satisfy the Fermi-Dirac distribution [15],
\begin{align}
n(\varepsilon_{\nu})=\dfrac{2}{h^{3}}\dfrac{1}{\exp\left({\dfrac{\varepsilon_{\nu}}{kT}}\right)+1}
\end{align}
where $h$ is Planck constant and $k$  is Boltzmann constant. With integrating the Eq.(5), the expression of deposition energy per unit time and unit volume is given by [15],
\begin{align}
\dfrac{\mathrm{d}E}{\mathrm{d}t\mathrm{d}V}=\dfrac{21\zeta(5)\pi^{4}}{h^{6}}KG_{F}^{2}F(r)(kT)^{9}
\end{align}
It is significant to derive the expression of $\dfrac{\mathrm{d}E}{\mathrm{d}t\mathrm{d}V}$ which is used to further the research on the converted energy rate on the different compact bodies [15]. The expression has something to do with the position, so has the temperature $T=T(r)$ called local temperature [15].

The local temperature $T(\mathbf{r})$ measured by a local observer is defined as $T(\mathbf{r})\sqrt{g_{00}(\mathbf{r})}=\text{constant}$ with $g_{00}$, a component of spacetime metric [15]. The neutrino temperature at the neutrinosphere can be explained as [15],
\begin{align}
T(r)\sqrt{g_{00}(r)}=T(R)\sqrt{g_{00}(R)}
\end{align}
where $R$ is the radius of a gravitational source. It is
elegant to replace the local temperature $T(r)$ in the future
calculation according to the identity (13). The luminosity
relating to the redshift can be selected as [15],
\begin{align}
L_{\infty}=g_{00}(R_{0})L(R_{0})
\end{align}
where the luminosity for a single neutrino species at the neutrinosphere is [15],
\begin{align}
L(R)=4\pi R_{0}^{2}\dfrac{7}{4}\dfrac{ac}{4}T^{4}(R)
\end{align}
where $a$ is the radiation constant and $c$ is the speed of light in vacuum. In order to replace the temperature associated with the observer's position, we combine Eq.(26), Eq.(27) and Eq.(28) and substitute them into Eq.(25) to obtain [15],
\begin{align}
\dfrac{\mathrm{d}E(\mathbf{r})}{\mathrm{d}t\mathrm{d}V}=\dfrac{21\zeta(5)\pi^{4}}{h^{6}}
KG_{F}^{2}k^{9}\left(\dfrac{7}{4}\pi ac\right)^{-\frac{9}{4}}L_{\infty}^{\frac{9}{4}}F(r)
\dfrac{\left[g_{00}(R)\right]^{\frac{9}{4}}}{\left[g_{00}(r)\right]^{\frac{9}{4}}}
R_{0}^{-\frac{9}{2}}
\end{align}
In addition to the radial coordinate, the metric components for the massive source surface also appear in the expression of deposition energy per unit time and unit volume like Eq.(29). We can calculate the radiation energy power in the background of the gravitational source by means of the deposition energy density over time as Eq.(29). In order to compute the angular integration $F(r)$ from Eq.(22), we should further the discussion on the variable $x$ in Eq.(23). We follow the procedure of Ref.[15] and solve the null geodesic in the spacetime of a spherically symmetric gravitational object [15] to show [20],
\begin{align}
x^{2}&=\sin^{2}\theta_{r}|_{\theta_{R}=0}\notag\\
&=1-\dfrac{R^{2}}{r^{2}}\dfrac{f(r)}{f(R)}\hspace{0.5cm}
\end{align}
Here $f(r)=g_{00}(r)$, a component of metric (2). It is useful to relate the variable $x=\sin\theta|_{\theta_{R}}=0$ to the environment structure around the gravitational object [20]. According to Eq.(22), the angular integration factor become the function of the metric. We can proceed the integration of rate per unit time and unit volume from Eq.(29) on the spherically symmetric volume around the gravitational source [21],
\begin{align}
\dot{Q}&=\dfrac{\mathrm{d}E}{\sqrt{g_{00}}\mathrm{d}t}\notag\hspace{10cm}\\
&=\dfrac{84\zeta(5)\pi^{5}}{h^{6}}KG_{F}^{2}k^{9}
\left(\dfrac{7}{4}\pi ac\right)^{-\frac{9}{4}}
L_{\infty}^{\frac{9}{4}}\left[g_{00}(R)\right]^{\frac{9}{4}}
R^{-\frac{9}{2}}\int_{R_{0}}^{\infty}\dfrac{r^{2}\sqrt{-g_{11}(r)}F(r)}
{g_{00}(r)}dr
\end{align}
where $g_{11}(r)=-\dfrac{1}{f(r)}$, also a component of the renormaliztion group improved Schwarzschild metric (2). The metrics of the curved spacetime have a considerable influence on the deposition energy rate. Here $\dot{Q}$ can reflect the total amount of energy converted from neutrinos to electron-positron pairs per unit time at any radius [15]. The conversion may become an explosion with the extremely large values of $\dot{Q}$. It is significant to compare the energy deposition rate (31) with the Newtonian quantities as [15, 20, 21],
\begin{align}
\dfrac{\dot{Q}}{\dot{Q}_{Newt}}=3\left[g_{00}(R)\right]^{\frac{9}{4}}
\int_{1}^{\infty}(x-1)^{4}\left(x^{2}+4x+5\right)\dfrac{y^{2}\sqrt{-g_{11}(Ry)}}
{g_{00}(Ry)^{\frac{9}{2}}}dy
\end{align}
with dimensionless variable $y=\dfrac{r}{R}$ and $g_{00}(r)$, $g_{11}(r)$ components of metric (1). According to the Ref.[15], we can present $\dfrac{\mathrm{d}\dot{Q}}{\mathrm{d}r}$ as the function of radial coordinate $r$ to exhibit the enhancement,
\begin{align}
\dfrac{\mathrm{d}\dot{Q}}{\mathrm{d}r}&=4\pi\left(\dfrac{dE}{dtdV}\right)\sqrt{-g_{11}(r)}r^{2}\hspace{9.5cm}\notag\\
&=\dfrac{168\zeta(5)\pi^{7}}{3h^{6}}KG_{F}^{2}k^{9}
\left(\dfrac{7}{4}\pi ac\right)^{-\frac{9}{4}}
L_{\infty}^{\frac{9}{4}}\notag\\
&\quad\times(x-1)^{4}\left(x^{2}+4x+5\right)
\left[\dfrac{g_{00}(R)}{g_{00}(r)}\right]^{\frac{9}{4}}R^{-\frac{5}{2}}
\sqrt{-g_{11}(r)}\left(\dfrac{r}{R}\right)^{2}
\end{align}
Here the derivative is a function of radial coordinate with the origin at the centre of
the gravitational source while involving the metric functions. It is necessary to wonder
how the structure of the compact body in asymptotic safety affect the neutrino annihilation
around one. The derivative function could tell us which kind of stars attracting
the annihilation could become source of gamma-ray burst.

It is important to discuss the ratio (32) in the quantum-gravity-corrected Schwarzschild
spacetime in the infrared limit. We can relate that $g_{00}(r)=f(r)$ and $g_{11}(r)=-\dfrac{1}{f(r)}$
for metric (1). The ratio (32) can be reformed as [15, 60-64],
\begin{align}
\dfrac{\dot{Q}}{\dot{Q}_{Newt}}=3\left[f(R)\right]^{\frac{9}{4}}\int_{1}^{\infty}
(x-1)^{4}\left(x^{2}+4x+5\right)\dfrac{y^{2}}{\left[f(Ry)\right]^{5}}\mathrm{d}y
\end{align}
where
\begin{align}
f(R)&=1-\dfrac{2M}{R}+\dfrac{2M\xi}{R^{3}}\\
f(Ry)&=1-\dfrac{2M}{R}\dfrac{1}{y}+\dfrac{2M\xi}{R^{3}}\dfrac{1}{y^{3}}
\end{align}
According to the metric function (2) and the running coupling (3), we rewrite the variable (30) as,
\begin{align}
x^{2}=1-\dfrac{1}{y^{2}}\dfrac{1-\dfrac{2M}{R}\dfrac{1}{y}
+\dfrac{2M\xi}{R^{3}}\dfrac{1}{y^{3}}}{1-\dfrac{2M}{R}+\dfrac{2M\xi}{R^{3}}}
\end{align}
\begin{figure}[t]
\setlength{\belowcaptionskip}{10pt} \centering
\includegraphics[width=12cm]{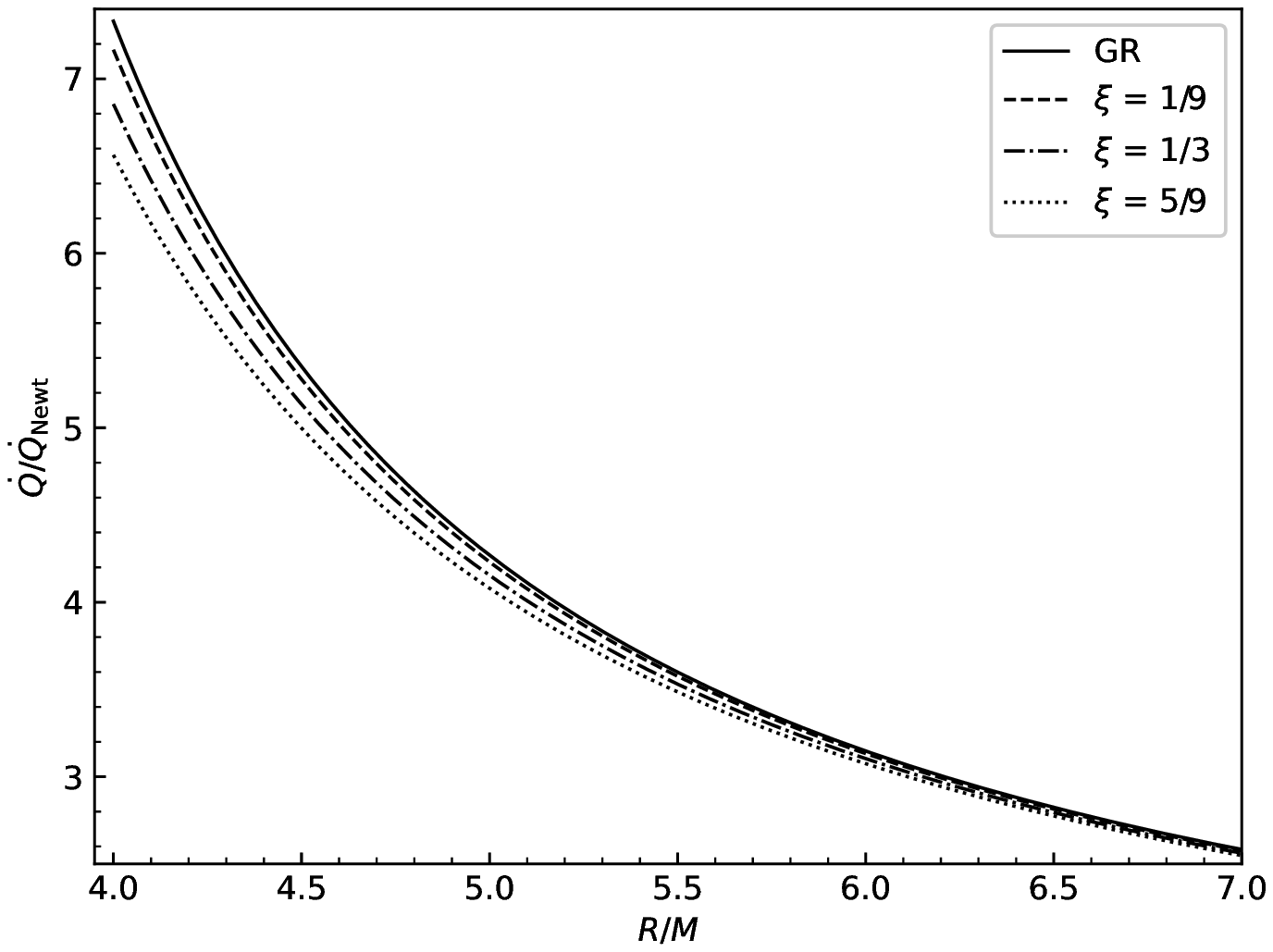}
\caption{The solid, dotted and dashed curves of the ratio $\dfrac{\dot{Q}}{\dot{Q_{Newt}}}$ as functions of the ratio $\dfrac{R}{M}$ for quantum-gravity factors $\xi=0, \dfrac{1}{9}, \dfrac{1}{3}, \dfrac{5}{9}$ respectively.}
\end{figure}
\begin{figure}[t]
\setlength{\belowcaptionskip}{10pt} \centering
\includegraphics[width=12cm]{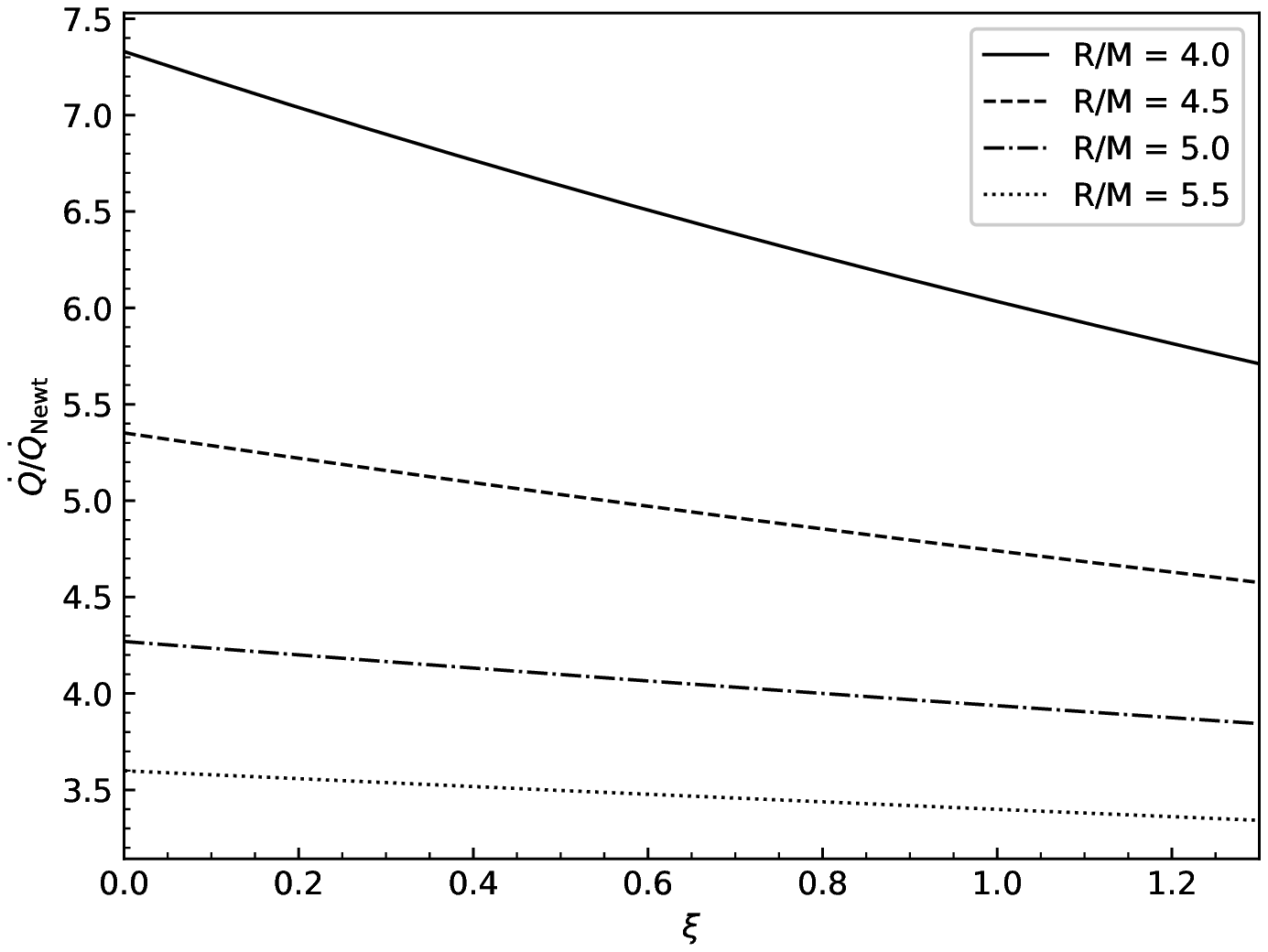}
\caption{The solid, dashed and dot-dashed curves of the ratio $\dfrac{\dot{Q}}{\dot{Q_{Newt}}}$ as functions of the quantum-gravity factors $\xi$ with $R=4M, 4.5M, 5M, 5.5M$ respectively.}
\end{figure}
In order to show the influence from quantum gravity, we should quantify the integral expression of ratio $\dfrac{\dot{Q}}{\dot{Q}_{Newt}}$ in Eq.(34) and depict the dependence on $\dfrac{R_{0}}{M}$ with the effect parameter $\xi$ in the Figures. Although the metric is corrected by the quantum gravity, it is difficult for us to estimate the influence on the energy deposition rate according to Eq.(34)-(36). We have to perform the complicated calculations to scrutinize the impact of neutrino annihilation associated with the spacetime structure around the compact bodies. It is necessary to gather data for the constraints over the values of $\xi$ [54]. During our plotting, we have to magnify itsvalues more than huge times in order to reveal the differences among the curves of the ratio $\dfrac{\dot{Q}}{\dot{Q}_{Newt}}$ for different magnitudes of factor $\xi$ from quantum-gravity effects on the neutrino pair annihilation around the black hole in asymptotic safety gravity because of extremely tiny values of $\xi$ from Ref.[54]. In Figure 1, the shape of ratio curves due to different values of parameter $\xi$ are similar. The influence from this kind of sources consist of two parts, mass $M$ and gravitational coupling $G(r)$ respectively. As a gravitational source, the values of ratio $\dfrac{\dot{Q}}{\dot{Q}_{Newt}}$ are much more than one, so this kind of stellar objects attracting the neutrino pair annihilation process may provide with a source of gamma-ray burst. For the sake of showing the asymptotic behaviour of ratio $\dfrac{\dot{Q}}{\dot{Q}_{Newt}}$ more clearly, we plot some curves to indicate that the ratio is decreasing function of $\xi$ from quantum corrections in Figure 2. In contrast to the case of accretion disk around a stellar object in asymptotic safety that the quantum gravity effect can enhance the disk's thermal properties [55], the more considerable influence will reduce the ratio $\dfrac{\dot{Q}}{\dot{Q}_{Newt}}$.

We also elaborate the derivative $\dfrac{\mathrm{d}\dot{Q}}{\mathrm{d}r}$ as a function of radius for several stellar masses denoted as $\dfrac{r}{R}$ with $R=3M$ to exhibit the promotion of $e^{-}e^{+}$ pair energy from the neutrino annihilation in Figure 3. It is similar that a slight larger variable $\xi$ from quantum gravity decreases the derivative $\dfrac{\mathrm{d}\dot{Q}}{\mathrm{d}r}$. The Figure 3 also indicate that the structure with smaller $\dfrac{r}{R}$ lead to the greater $\dfrac{\mathrm{d}\dot{Q}}{\mathrm{d}r}$. It is interesting that the increase is much larger near the surface of the neutron star similar to the case of Ref.[15] although the compact source surrounded by the annihilation includes the quantum gravity influence.

\begin{figure}[t]
\setlength{\belowcaptionskip}{10pt} \centering
\includegraphics[width=12cm]{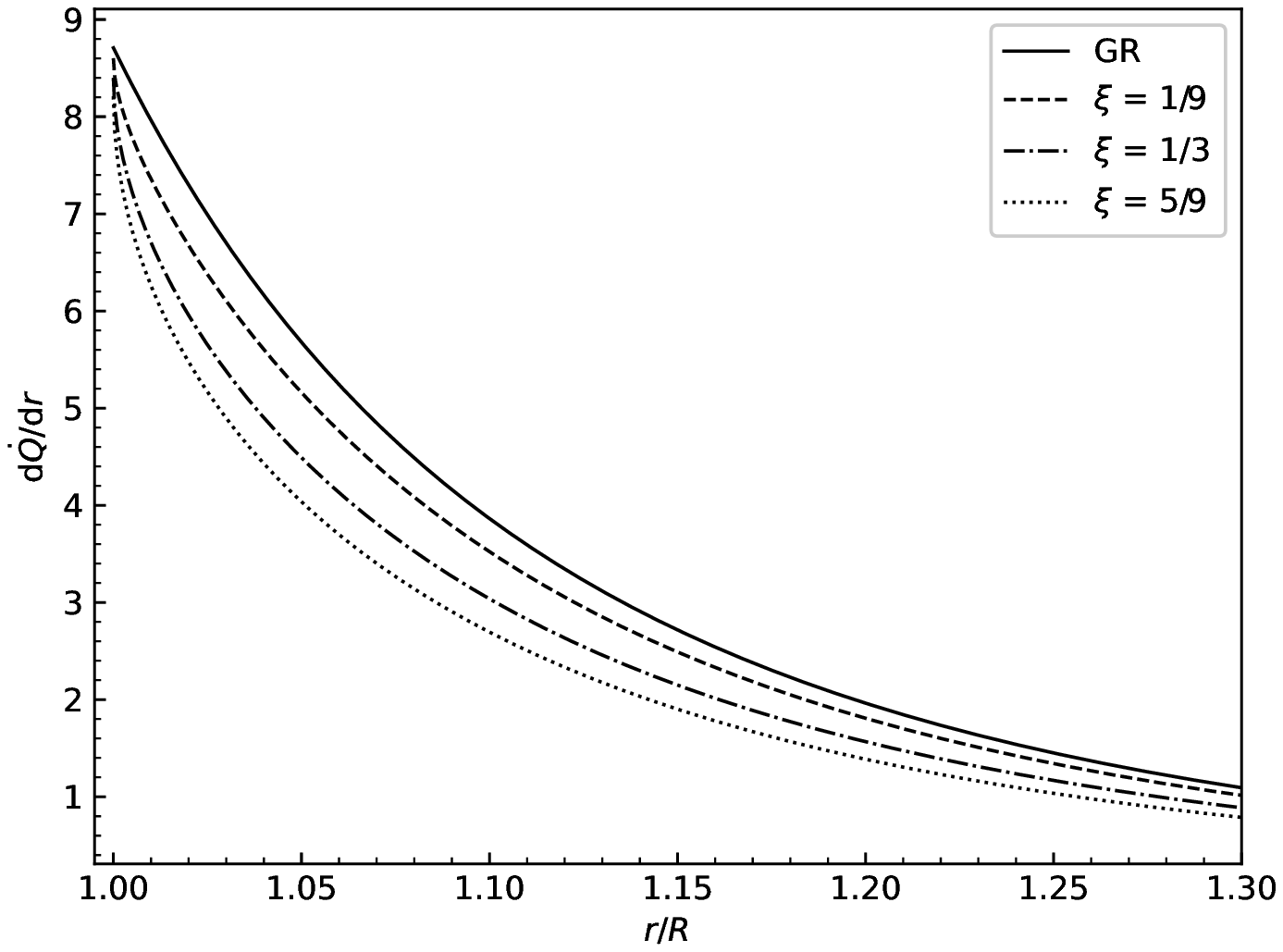}
\caption{The solid, dashed, dot-dashed and dotted curves of the derivative $\dfrac{\mathrm{d}\dot{Q}}{\mathrm{d}r}$ as functions of the ratio $\dfrac{r}{R}$ with $R=3M$ for quantum-gravity factors $\xi=0, \dfrac{1}{9}, \dfrac{1}{3}, \dfrac{5}{9}$ respectively.}
\end{figure}

\section{Conclusion}

The neutrino pair annihilation $\nu\bar{\nu}\longrightarrow e^{-}e^{+}$ around a Schwarzschild black hole in asymptotic safety is discussed. The quantum gravity influence on the black holes can be explored in different directions. We derivative and calculate the emitted energy rate ratio as $\dfrac{\dot{Q}}{\dot{Q}_{Newt}}$ and the emitted energy rate $\dot{Q}$ analyzed in the background of quantum-gravity corrected massive source. It is found that the quantum-gravity effect does not advance the ratio $\dfrac{\dot{Q}}{\dot{Q}_{Newt}}$, but damps $\dot{Q}$, the energy released per unit time, in contrast to the case of accretion disk under the same conditions. The larger $\xi$, the parameter encoding the quantum gravity effects on the spacetime structure, will reduce the energy deposition rate. The ratio $\dfrac{\dot{Q}}{\dot{Q}_{Newt}}$ shows numerically that the annihilation $\nu+\bar{\nu}\longrightarrow e^{-}+e^{+}$ around the quantum-gravity corrected black hole can proceed as a source for the gamma-ray burst. This kind of annihilation can occur around the Type-II supernova or collapsing neutron stars. Within the region of $\xi$, this kind of annihilation process could be proposed as a well-qualified candidate of gamma-ray bursts.

\section*{Acknowledge}

This work is partly supported by the Shanghai Key Laboratory of Astrophysics 18DZ2271600.


\begin{thebibliography}{99}
\bibitem{Eichler}D. Eichler, M. Livio, T. Piran, D. N. Schramm,
Nature 340(1989)126
\bibitem{Paczynski}B. Paczynski, Astrophys. J. 363(1990)218
\bibitem{Meszaros}P. Meszaros, M. J. Rees, MNRAS 257(1992)29
\bibitem{Woosley}S. E. Woosley, Astrophys. J. 405(1993)273
\bibitem{Mochkovitch}R. Mochkovitch, M. Hernanz, J. Isern, X.
Martin, Nature 361(1993)236
\bibitem{Davies}M. B. Davies, W. Benz, T. Piran, F. K. Thieleman,
Astrophys. J. 431(1994)742
\bibitem{Jaroszynski}M. Jaroszynski, A A, 305(1996)839
\bibitem{Janka}H. Janka, M. Ruffert, A A, 307(1996)L33
\bibitem{Ruffert}M. Ruffert, H. Janka, K. Takahashi, G. Schafer, A
A, 319(1997)122
\bibitem{Ruffert}M. Ruffert, H. Janka, A A, 338(1998)535
\bibitem{Popham}R. Popham, S. E. Woosley, C. Fryer, Astrophys. J.
518(1999)356
\bibitem{MacFadyen}A. MacFadyen, S. E. Woosley, Astrophys. J.
524(1999)262
\bibitem{Ruffert}M. Ruffert, H. Janka, A A, 344(1999)573
\bibitem{Aloy}M. A. Aloy, E. Muller, J. M. Ibanez, J. M. Martin, A.
MacFadyen, Astrophys. J. 531(2000)L119
\bibitem{Salmonson}J. D. Salmonson, J. R. Wilson, Astrophys. J.
517(1999)859
\bibitem{Asano}K. Asano, T. Fukuyama, Astrophys. J. 531(2000)949
\bibitem{Asano}K. Asano, T. Fukuyama, Astrophys. J. 546(2001)1019
\bibitem{Miller}W. A. Miller, N. D. George, A. Kheyfets, J. M.
McGhee, Astrophys. J. 583(2003)833
\bibitem{Birkl}R. Birkl, M. A. Aloy, H. Janka, E. Muller, A A
463(2007)51
\bibitem{Lambiase}G. Lambiase, L. Mastrototaro, Astrophys. J.
904(2020)1\\
A. Prasanna, S. Goswami, Phys. Lett. B526(2002)27
\bibitem{Lambiase}G. Lambiase, L. Mastrototaro, Eur. Phys. J.
C81(2021)932
\bibitem{Shi}Y. Shi, H. Cheng, EPL 140(2022)49001
\bibitem{Poddar}T. K. Poddar, S. Goswai, A. K. Mishra, Eur. Phys.
J. C83(2023)223
\bibitem{Weinberg}S. Weinberg, Gravitation and Cosmology:
Principles and Applications of the General Theory of Relativity,
John Wiley Sons Inc., 1972\\
M. Carmeli, Classical Fields: General Relativity and Gauge Theory,
John Wiley Sons, Inc., 1982
\bibitem{Chandrasekhar}S. Chandrasekhar, The Mathematical Theory
of Blck Holes, Oxford University Press, Inc., 1992
\bibitem{Ashtekar}A. Ashtekar, Phys. Rev. Lett. 57(1986)2244
\bibitem{Ashtekar}A. Ashtekar, Phys. Rev. D36(1987)1587
\bibitem{Rovelli}C. Rovelli, Living Rev. Relativ. 1(1998)1
\bibitem{Ashtekar}A. Ashtekar, J. Baez, A. Corichi, K. Krasnov,
Phys. Rev. Lett. 80(1998)904
\bibitem{Perez}A. Perez, Class. Quantum Grav. 20(2003)R43
\bibitem{Horava}P. Horava, Phys. Rev. D79(2009)084008
\bibitem{Horava}P. Horava, Phys. Rev. Lett. 102(2009)161301
\bibitem{Kiritsis}E. Kiritsis, G. Kofinas, Nucl. Phys.
B821(2009)467
\bibitem{Cognola}G. Cognola, R. Myrzakulov, L. Sebastiani, S.
Vagnozzi, S. Zerbini, Class. Quantum Grav. 33(2016)225014
\bibitem{Reuter}M. Reuter, Phys. Rev. D57(1998)971
\bibitem{Bonanno}A. Bonanno, M. Reuter, Phys. Rev. D62(2000)043008
\bibitem{Reuter}M. Reuter, E. Tuiran, Phys. Rev. D83(2011)044041
\bibitem{Koch}B. Koch, F. Saueressig, Class. Quantum Grav.
31(2014)015006
\bibitem{Pawlowski}J. M. Pawlowski, D. Stock, Phys. Rev.
D98(2018)106008
\bibitem{Gonzalez}C. Gonzalez, B. Koch, Int. J. Mod. Phys.
A31(2016)1650141
\bibitem{Casadio}R. Casadio, S. D. H. Hsu, B. Mirza, Phys. Lett.
B695(2011)317
\bibitem{Kofinas}G. Kofinas, V. Zarikas,  JCAP 1510(2015)069
\bibitem{Bonano}A. Bonano, B. Koch, A. Platania, Found. Phys.
48(2018)1393
\bibitem{Bosma}L. Bosma, B. Knorr, F. Sauressig, Phys. Rev. Lett.
123(2019)101301
\bibitem{Falls}K. Falls, D. F. Litim, Phys. Rev. D89(2014)084002
\bibitem{Falls}K. Falls, Int. J. Mod. Phys. A27(2012)1250019
\bibitem{Ward}B. F. L. Ward, Acta Phys. Polon. B37(2006)1967
\bibitem{Yang}R. Yang, Phys. Rev. D92(2015)084011
\bibitem{Farooq}M. U. Farooq, A. K. Ahmed, R. Yang, M. Jamil,
Chin. Phys. C44(2020)065102
\bibitem{Zuluaga}F. H. Zuluaga, L. A. Sanchez, Chin. Phys.
C45(2021)075102
\bibitem{Held}A. Held, R. Gold, A. Eichhorn, JCAP 1906(2019)029
\bibitem{Lu}X. Lu, Y. Xie, Eur. Phys. J. C79(2019)1016
\bibitem{Kumar}R. Kumar, B. P. Singh, S. G. Ghosh, Ann. Phys.
420(2020)168252\\
A. Eichhorn, A. Held, JCAP 2105(2021)073
\bibitem{Lambiase}G. Lambiase, F. Scardigli, Phys. Rev.
D105(2022)124054
\bibitem{Zuluaga}F. H. Zuluaga, L. A. Sanchez, Eur. Phys. J.
C81(2021)840
\bibitem{Haroon}S. Haroon, M. P. Pavlovic, M. Sossich, A. Wang,
Eur. Phys. J. C78(2018)519
\bibitem{Page}D. N. Page, K. S. Thorne, Astrophys. J. 191(1974)499
\bibitem{Zhou}B. Zhou, A. Abdikamalov, D. Ayzenberg, C. Bambi, S.
Nampalliwar, A. Tripathi, JCAP 2101(2021)047\\
S. Vagnozzi, R. Roy, Y. Tsai, L. Visinelli, M. Afrin, A.
Allahyari, P. Bambhaniya, D. Dey, S. G. Ghosh, P. S. Joshi, K.
Jusufi, M. Khodadi, R. K. Walia, A. Ovgun, C. Bambi,
"Horizon-scale tests of gravity theories and fundamental physics
from the Event Horizon Telescope image of Sagittarius $A^{\ast}$",
arXiv: 2205.07787
\bibitem{Zhang}Y. Zhang, M. Zhou, C. Bambi, Eur. Phys. J.
C78(2018)376
\bibitem{Benedetti}D. Benedetti, P. F. Machado, F. Saueressig,
Mod. Phys. Lett. A24(2009)2233
\bibitem{Cai}Y. Cai, D. A. Easson, JCAP 1009(2010)002
\bibitem{Falls}K. Falls, C. R. King, D. F. Litim, K.
Nikolakopoulos, C. Rahmede, Phys. Rev. D97(2018)086006
\bibitem{Falls}K. Falls, D. F. Litim, J. Schroder, Phys. Rev.
D99(2019)126015
\bibitem{Kluth}Y. Kluth, D. F. Litim, arXiv: 2008.09181
\bibitem{Cardoso}V. Cardoso, V. F. Foit, M. Kleban, JCAP
1908(2019)006
\bibitem{Li}Z. Li, Y. Piao, Phys. Rev. D100(2019)044023
\bibitem{Bronnikov}K. A. Bronnikov, R. A. Konoplya, Phys. Rev.
D101(2020)064004
\bibitem{Tsukamoto}N. Tsukamoto, T. Kokubu, Phys. Rev.
D101(2020)044030
\bibitem{Nambu}Y. Nambu, S. Nota, Y. Sakai, Phys. Rev.
D100(2019)064037




\end{thebibliography}
\end{document}